# Single-peak and narrow-band mid-infrared thermal emitters driven by mirror-coupled plasmonic quasi-BIC metasurfaces

Sen Yang[1,3,‡], Mingze He[2,‡], Chuchuan Hong[1], Josh Nordlander[4], Jon-Paul Maria[4], Joshua D. Caldwell[2,3,5], and Justus C. Ndukaife[1,3*]

[1]Department of Electrical and Computer Engineering, Vanderbilt University, Nashville, TN, USA 37235

[2]Department of Mechanical Engineering, Vanderbilt University, Nashville, TN, USA 37235

[3]Interdisciplinary Materials Science Program, Vanderbilt University, Nashville, TN, USA 37240

[4]Department of Materials Science and Engineering, The Pennsylvania State University, University Park, Pennsylvania, USA 16802

[5]Sensorium Technological Labs, 6714 Duquaine Ct, Nashville, TN, USA 37205

[‡]These authors contribute equally to this work.

*Correspondence: justus.ndukaife@vanderbilt.edu

ABSTRACT GRAPH

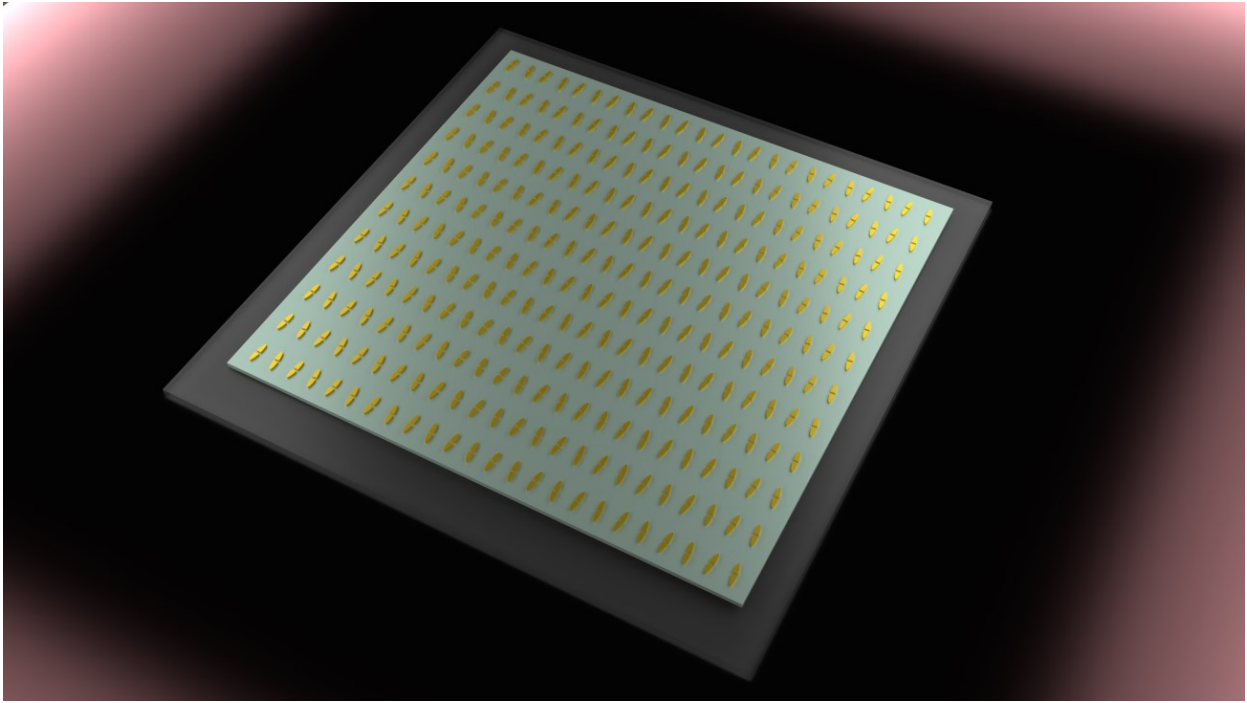

ABSTRACT Wavelength-selective thermal emitters (WS-EMs) hold considerable appeal due to the scarcity of cost-effective, narrow-band sources in the mid-to-long-wave infrared spectrum. WS-EMs achieved via dielectric materials typically exhibit thermal emission peaks with high quality factors (Q factors), but their optical responses are prone to temperature fluctuations. Metallic EMs, on the other hand, show negligible drifts with temperature changes, but their Q factors usually hover around 10. In this study, we introduce and experimentally verify a novel EM grounded in plasmonic quasi-bound states in the continuum (BICs) within a mirror-coupled system. Our design numerically delivers an ultra-narrowband single peak with a Q factor of approximately 64, and near-unity absorptance that can be freely tuned within an expansive band of more than 10 µm. By introducing air slots symmetrically, the Q factor can be further augmented to around 100. Multipolar analysis and phase diagrams are presented to elucidate the operational principle. Importantly, our infrared spectral measurements affirm the remarkable resilience of our designs' resonance frequency in the face of temperature fluctuations over 300°C. Additionally, we develop an effective impedance model based on the optical nanoantenna theory to understand how further tuning of the emission properties is achieved through precise engineering of the slot. This research thus heralds the potential of applying plasmonic quasi-BICs in designing ultra-narrowband, temperature-stable thermal emitters in mid-infrared. Moreover, such a concept may be adaptable to other frequency ranges, such as near-infrared, Terahertz, and Gigahertz.

## 1. Introduction

The creation of affordable and efficient mid-infrared narrowband light sources is highly sought after for a wide array of applications[1]. These include, but are not limited to, free-space communications[2], infrared beacons, bar-codes, and enhancements in monitoring environmental pollutants and toxins through molecular sensing methodologies, such as non-dispersive infrared (NDIR) sensing[3–5]. According to Kirchhoff's law and Planck's law, a narrowband absorber could serve as a narrowband thermal emitter (EM) at elevated temperatures, since the emissivity equals the absorptivity for a given frequency, direction, polarization, and temperature for all reciprocal systems. In simple terms, an efficient absorber also serves as an efficient emitter. In a seminal demonstration in 2002, Greffet et al.[6] showed that a silicon carbide grating could emit thermal light with enhanced spatial coherence compared to an unpatterned surface. Since then, capabilities for producing correlated electromagnetic fields in the far field by carefully engineering the near-field details of nanostructures have been significantly expanded. Numerous types of EMs operating in the mid-infrared have been reported, including gratings[7,8], metamaterials[9,10], metasurfaces[11–14], photonic crystals[15,16], and stacked films[17–21]. Among these, designs based on dielectric materials often exhibit a high Q factor[22] (Q ~ $10^1$ - $10^2$) due to their low intrinsic losses, offering great potential for narrow-band applications. However, their optical responses are typically temperature-dependent[23], as the refractive index of dielectric materials (like silicon) significantly varies with temperature[24,25], causing spectral drift in thermal emission. Conversely, metallic EMs demonstrate relatively stable performance across a wide temperature range due to the dominance of the imaginary part of the refractive index of the metal in the mid- to long-wave infrared[9]. However, the inherent (Ohmic) losses in metals significantly broadens the emission peak (Q ~ 10, see Table 1). In response to these challenges, in this work, we leverage the high-Q properties of bound states in the continuum (BICs).

The concept of BICs has been recently introduced to plasmonics to achieve high-Q spectral resonances[26–32]. BICs, initially proposed in quantum mechanics by von Neumann and Eugene Wigner in 1929[33], have quickly emerged as an effective strategy for achieving the high Q factor and strong field enhancement in photonic systems for diverse applications[34,35]., including lasing[36], sensing[37], and nonlinear effects[38]. BICs are eigenmodes embedded above the light cone but uncoupled from the radiation continuum of free space[39]. A true BIC has an infinite quality factor and vanishing resonance linewidth and thus not accessible. When a true BIC is perturbed, it is transformed to a quasi-BIC with a large finite Q factor, highly enhanced electromagnetic field intensity enhancement, and a measurable resonance linewidth. Investigations of BICs in plasmonic structures remain challenging primarily due to their intrinsic losses, with most attention currently centered on the symmetry-protected type[32], resulting from the incompatibility of the mode's spatial symmetry with the symmetry of the outgoing radiating wave. Several early works reported plasmonic BICs based on grating structures[26,28,30,32] or 3D laser nanoprinting of plasmonic nanofins[27,29,31], while ellipse-shaped BICs[40,41] in plasmonics have been very recently reported for applications in optical modulation[42] and absorbance spectroscopy[43]. In plasmonic BIC systems, the radiation losses into free space are entirely eliminated at the BIC point, meaning the total Q factor is solely constrained by the dissipative losses and can reach a high value[26,29] ($Q \sim 10^2$), rendering them excellent candidates for narrow-band EMs. Yet, the benefits of plasmonic quasi-BICs, e.g., high quality factors, have not yet been exploited in the field of narrowband EM designs. Moreover, it is possible to realize temperature-stable structures with plasmonic materials[9].

In this work, we introduce a single-peak, narrow-band thermal emitter operating in the mid-infrared region, premised on the MIM configuration powered by plasmonic quasi-BICs. Our plasmonic metasurface comprises elliptical gold nanoresonators arranged in a zigzag pattern[40], as shown in Fig. 1a and b. This design numerically yields an ultra-narrowband absorptance peak with a Q factor of approximately 64, and a near-unity absorptance of 0.97 (Fig. 1c). Achieving such a high Q factor and near-unity absorptance simultaneously is unattainable in conventional metallic designs. By incorporating symmetrical slots (Fig. 1d and e), we are able to boost the Q factor while keeping a reasonably high absorptance. Through numerical calculations, we further explore the multipolar character of the induced plasmonic quasi-BIC modes and the angular dispersion of thermal emission. We validate the performance of our designs using Fourier-transform infrared (FTIR) spectroscopy. The results confirm the resonance frequency of our designs to be resilient to temperature fluctuations, exhibiting only minor peak broadening at elevated temperatures. Moreover, we present an effective impedance model based on optical nanoantenna theory. From this model, we qualitatively elucidate how broad tuning of the emission properties is achieved through engineering the geometry of the slots observed in our experiments. Compared to other strategies like Tamm plasmon polariton thermal emitters[20,44], phonon polariton resonances in polar materials[45,46], and photonic crystals[47], our design strikes a balance between performance, design tunability and fabrication complexity. It facilitates a tunable single-frequency peak ranging from 3 μm to 13 μm, while maintaining a comparatively straightforward fabrication process. Our presented methodology can be readily extended to operation at high temperatures by leveraging refractory materials[48] like tantalum[49] and tungsten[50], or transition metal nitrides like titanium nitride[51,52], thus opening the way for using plasmonic quasi-BICs for high-power EMs. By engineering the BIC structures, other controls including polarization, angular dispersion, and even far-field profiles may also be achieved[53,54]. Such concepts are also readily adaptable to other frequency ranges, such as near-infrared, terahertz, and microwave regimes.

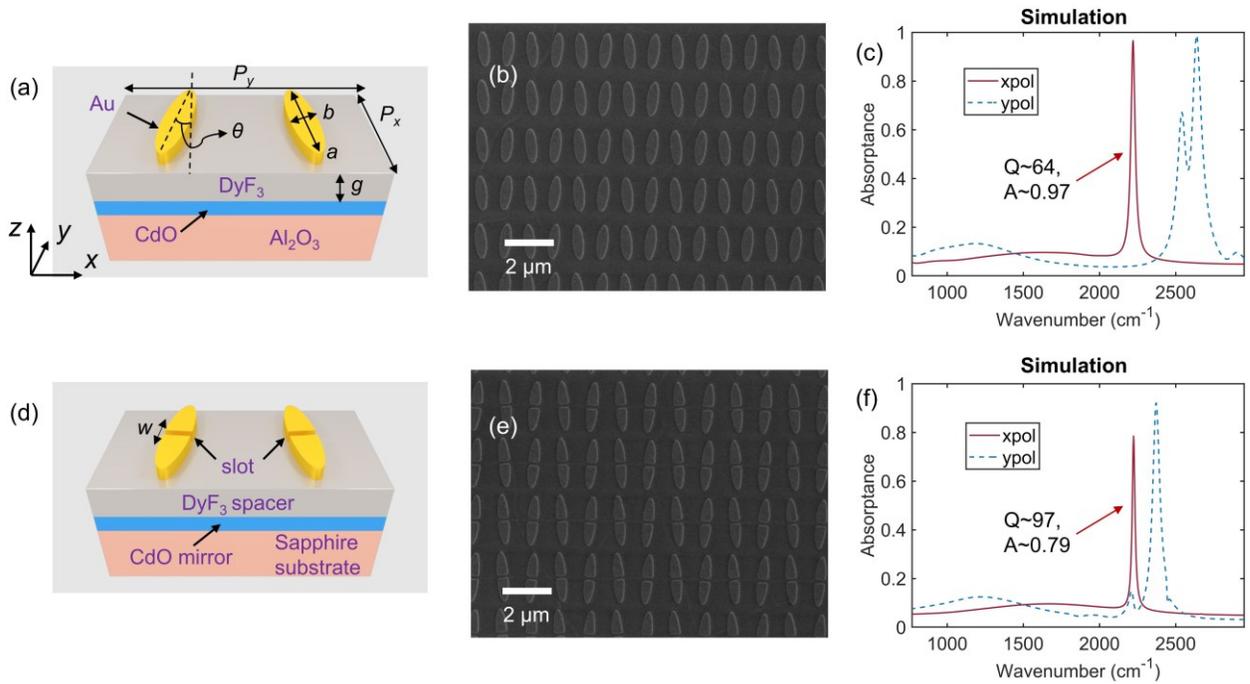

**Figure 1.** Design and spectra simulations of the thermal emitters. (a) Schematic illustration of a unit cell of the "0 slots" metasurface on a DyF$_3$ spacer layer, followed by a CdO mirror and a sapphire substrate. The metasurface comprises square lattices (periods, $P_x = P_y = 2100$ nm) of dimers formed by two gold elliptical resonators with a tilted angle of $\theta = 6°$. The geometrical parameters: $a = 1400$ nm, $b = 420$ nm, $g = 650$ nm. The thickness of the gold resonators and the CdO mirror is $H$ = 200 nm and 400 nm, respectively. The incident light is linearly polarized plane wave. (b) Representative scanning electron microscope (SEM) image of a "0 slots" sample. (c) Simulated spectra of the "0 slots" configuration for $x$ and $y$ polarizations

under normal incidence over 3 μm to 13 μm, respectively. The primary quasi-BIC peak (red solid curve) shows a Q factor of ~ 64 and absorptance of ~ 0.97. (d) Schematic illustration of a unit cell of the "1 slot" configuration. The slot direction is aligned with the short axis of the ellipse to maintain the structure symmetry. The geometrical parameters: $P_x = P_y = 2508$ nm, $\theta = 5°$, $a = 2006$ nm, $b = 401$ nm, $g = 650$ nm, $w = 100$ nm. The thickness of the gold resonators and the CdO mirror remains the same. (e) Representative SEM image of a "1 slot" sample. (f) Simulated spectra of the "1 slot" configuration for *x* and *y* polarizations under normal incidence over 3 μm to 13 μm, respectively. The primary quasi-BIC peak (red solid curve) shows a Q factor of ~ 97 and absorptance of ~ 0.79.

## 2. Main

In accordance with Kirchhoff's law, achieving a highly efficient EM necessitates near-unity absorptance, essentially creating a perfect absorber (PA). The success of an efficient plasmonic PA heavily relies on the careful management and alignment of radiative and dissipative losses of the system[55]. In such a system, the total Q factor ($Q_{tot}$) that is shown in the spectrum is governed by the radiative Q factor ($Q_{rad}$) and the dissipative Q factor ($Q_{dis}$) with their relationship being expressed as $Q_{tot} = (Q_{rad}^{-1} + Q_{dis}^{-1})^{-1}$. To achieve near-unity absorptance, precise control is needed to ensure $Q_{rad} = Q_{dis}$, denoting the critical coupling condition (CC condition) [56,57]. And a high Q factor demands a concurrent elevation of both $Q_{rad}$ and $Q_{dis}$. A common approach in designing plasmonic PAs as EMs relies on the metal-insulator-metal (MIM) configuration[58]. However, most exhibited MIM-configured EMs show a comparatively low Q factor, usually under 10 (see Table 1). Even though plasmonic EMs in mid-infrared supporting surface lattice resonances (SLRs) can achieve a fairly high Q factor (for instance, Q ~ 32 reported by Pusch et al.[59]), these modes typically necessitate an up-down mirror symmetry (that is, embedding in a homogeneous medium), and their multiple-peak nature restricts their ability to achieve a single peak within a wide frequency range[60]. Instead, here we propose an EM that is composed of a plasmonic quasi-BIC metasurface sitting on a $DyF_3$ dielectric spacer layer, separated by a CdO metallic reflector/mirror from a sapphire substrate, featuring a typical MIM configuration (see

Figure 1a). The plasmonic metasurface consists of elliptical gold resonators with a tilted angle arranged in a zigzag array, as shown by the scanning electron microscope (SEM) image in Figure 1b. We specifically select these materials to ensure that the optical responses of the emitter remain stable under different temperatures. Our proposed system is designed to support a resonance at 2222 cm$^{-1}$, which overlaps with one absorption peak of $N_2O$ falling within our targeted frequency range[61], since one of our intended applications is NDIR gas sensing. We carefully engineer the critical coupling between the intrinsic material losses and the radiative losses of the quasi-BIC mode (see Figures 2f and g), resulting in an ultra-narrowband absorptance peak with a calculated Q factor of approximately 64 and near-unity absorptance of 0.97 under *x*-polarized incidence, as highlighted in Figure 1c. Moreover, the quasi-BIC mode can be generated over a substantially wide band (3 μm to 13 μm), resulting in a single peak that can be tuned within most of the mid-infrared window. The position of the peak can be easily and flexibly adjusted by altering the geometry of the quasi-BIC mode[40,43]. Given that the metasurface exhibits mirror symmetry along the *y*-axis, but not the *x*-axis, the optical responses corresponding to the quasi-BIC mode vary between the *x* and *y* polarizations. By carefully adjusting the geometric parameters, we intentionally shift the resonance under *y*-polarized incidence, distancing it from the primary quasi-BIC peak, as demonstrated in Figure 1c. Such a design may prove useful in applications that necessitate dual-band functionality[10]. Next, we incorporate an air slot into each of the elliptical resonators. The slots, symmetrically positioned, extend through each gold resonator to ensure the overall symmetry is preserved, as shown in Figure 1d and e. To be concise, we refer to this slotted system as the "1 slot" configuration, and similarly, we have "0 slots" for the original system. This modification allows for an even higher Q factor of approximately 97, while maintaining a peak absorptance of 0.79 that is comparable with many demonstrations[6,16,20,45]. It is worth noting that this innovative slotted design offers an alternative method for enhancing the resonance of quasi-BIC metasurfaces, apart from current techniques such as engineering the structure symmetry from individual resonators (e.g, the tilted angle ($\theta$) in this work) or altering the dimensions of elements to excite supercavity modes[62]. This approach has been validated in our recent study involving an all-dielectric quasi-BIC metasurface[63]. More comprehensive details on how the slot geometry can influence this plasmonic quasi-BIC mode can be found in Figure 5.

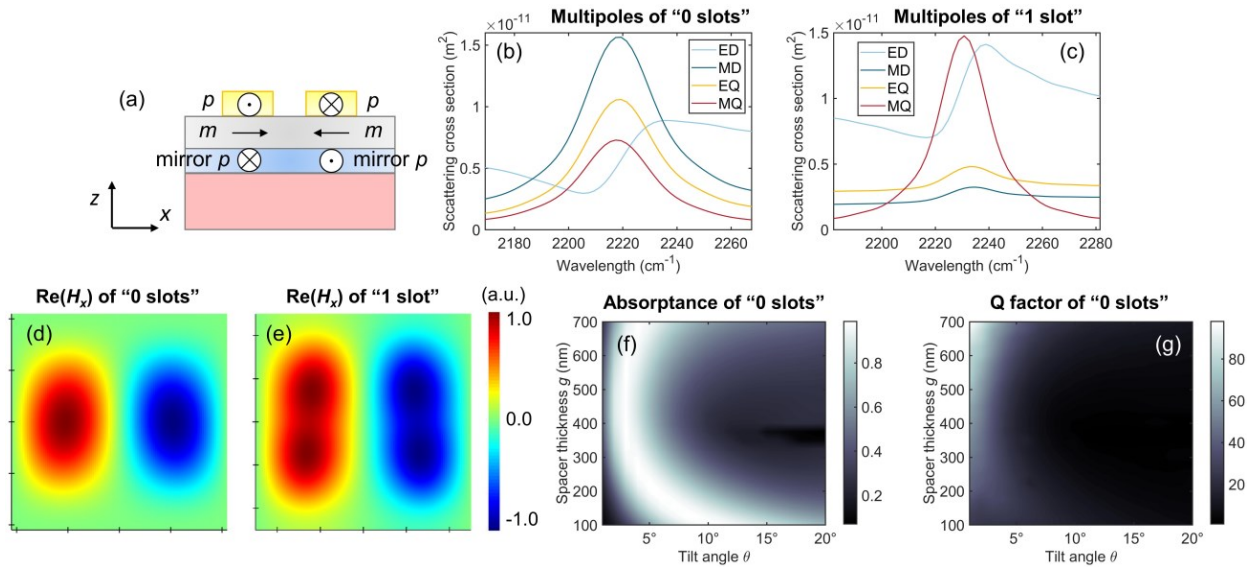

**Figure 2.** Working principle and design optimization. (a) Side-view schematic illustration depicting the displacement current flows and the induced magnetic dipole moments. p, electric dipole; m, magnetic dipole. (b) Multipole decomposition results for the quasi-BIC resonance corresponding to the "0 slots" configuration and (c) for the "1 slot" configuration. ED, electric dipole; MD, magnetic dipole; EQ, electric quadrupole; MQ, magnetic quadrupole. The multipole decomposition analysis was applied to the whole unit cell and contained electric field information contributed by the elliptical resonators, the spacer layer as well as the reflector layer. (d) Distributions of the *x* component of the magnetic field (*xy* plane) at the middle of the spacer layer on quasi-BIC resonance corresponding to the "0 slots" configuration and (e) for the "1 slot" configuration. (f) 2D phase diagrams showing the absorptance and (g) total Q factor calculated from numerical simulations as a function of the spacer thickness (*g*) and tilted angle ($\theta$).

To comprehend the generation of the quasi-BIC modes in this mirror-coupled plasmonic system, we examine the pair of inverted circulating displacement currents induced along the ellipse's long axis on each tilted gold resonator within a unit cell (see Figure 2a). This inverted behavior is a characteristic trait found in the electric dipole moments of similar tilted-ellipse-based all-dielectric quasi-BIC metasurfaces[40,64]. With the presence of the metallic reflector layer, an anti-parallel mirror current is induced beneath each tilted gold resonator, forming a loop and generating a magnetic dipole (MD) moment in the spacer layer. Intriguingly, the two induced MD moments from each ellipse on the reflector also exhibit an opposite orientation within the unit cell. The presence of this unique mode suggests that the inherent characteristics of the tilted-ellipse-based, symmetry-protected quasi-BIC mode are preserved in this MIM system. The multipole decomposition results (see Section S1 of the SI for more details[65–68]) depicted in Figure 2b highlight the dominance of the $z$ component of the MD component on quasi-BIC resonance, which presents the characteristic features common to the tilted-ellipse-based all-dielectric quasi-BIC metasurfaces[65]. The distribution of the $x$ component of the magnetic field in Figure 2d distinctly reveals a pair of inverted MD moments. Upon the introduction of the air slot, the circulating displacement current induced on the tilted gold resonator is disrupted and divided into two separate streams, as indicated in Figure S3 of the SI. Consequently, two parallel MD moments are generated in the spacer layer beneath each resonator, as illustrated in Figure 2e. This is further corroborated by the substantial increase in contributions from the magnetic quadrupole (MQ) component on quasi-BIC resonance, as depicted in Figure 2c. As underscored in our previous research[63], the symmetric incorporation of air slots into the electric dipole resonators can considerably augment both the Q factor and the electric field, relative to the original "0 slots" configuration, within such an ellipse-based quasi-BIC system. This assertion is corroborated by the narrower linewidth of the MQ component observed in Figure 2c, as well as the intensified electric field enhancement exhibited in Figure S3 of the SI. The amplification of the total Q factor ($Q_{tot}$) results in a deviation from the CC condition and consequently a decrease in absorptance.

Building on the discussions by Wang et al.[43], we construct phase diagrams[55] by adjusting the spacer thickness (*g*) and tilted angle (*θ*), to arrive at an optimal design for the "0 slots" EM – one that yields a high Q factor and near-unity absorptance, as demonstrated in Figures 2f and g. Figure 2g shows a general increase in the total Q factor ($Q_{tot}$) as the spacer becomes thicker, primarily due to the reduction in dissipative losses. Taking into account our fabrication capabilities, we select *g* = 650 nm and identify the optimal tilted angle at *θ* = 6° from the CC condition depicted in Figure 2f. Additionally, the inclusion of the air slot offers more flexibility in manipulating both the radiative and dissipative losses. Consequently, by sweeping the geometrical parameter space inclusive of the slot width (*w*), one may further enhance the performance of the "1 slot" configuration reported in this work.

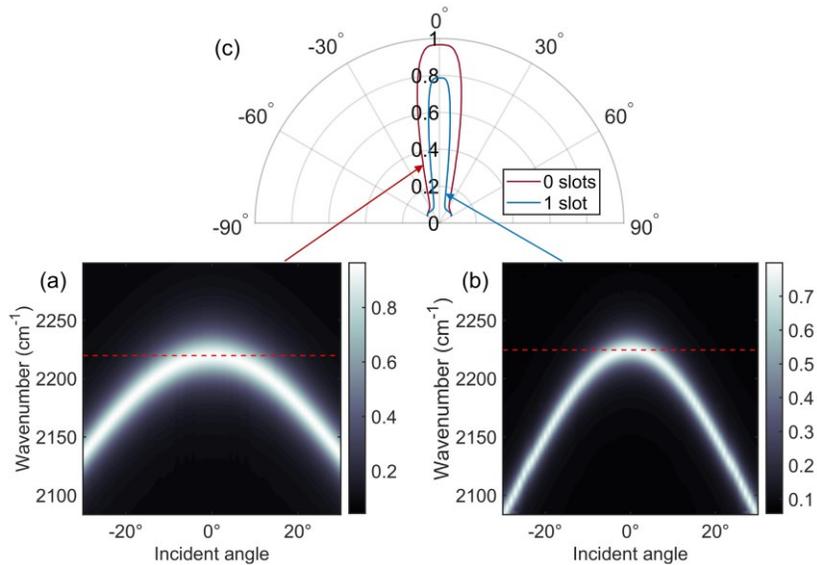

**Figure 3.** Characterization of angular dispersion. (a) Band diagrams showing the absorptance spectra as a function of incident angles for the "0 slots" configuration and (b) for the "1 slot" configuration. The red dash line denotes the frequency of the band-edge mode. (c) Polar diagram comparing the angular dispersion properties of absorption (i.e., emission) for the "0 slots" and "1 slot" configurations.

The directivity of the emission serves as a crucial characteristic of an EM. To quantify the directivity of the thermal emitter, we use the spatial coherence formula, $L_c \approx \sqrt{b\tau}$, where $b$ represents the band curvature near the band-edge and $\tau$ signifies the mode lifetime[54]. A larger spatial coherence can be beneficial for applications such as wavefront engineering[1], e.g., lensing[54] and polarization control[53], while a small band curvature (small $L_c$) is preferred when light at a range of angles is collected, leading to reduced spectral broadening. To analyze the directionality, we numerically construct the band diagrams for the TM-polarized excitation (where both the wave vector and polarization align along the *xz* plane; see Figure S5 of the SI for details) for both "0 slots" and "1 slot" configurations, as displayed in Figures 3a and b. Notably, the "0 slots" configuration exhibits a flatter band compared to the "1 slot" configuration, indicating that the thermal emission from the "0 slots" design possesses less directivity. This is further confirmed by creating the polar diagram for the band-edge modes, as depicted in Figure 3c. For the "1 slot" design, the amplified Q factor and the sharper band enhance the distance the quasi-BIC mode traverses in-plane before scattering. Consequently, this increases the linewidth of the spectral degree of coherence[6,54] from $2\lambda_0$ to $3.6\lambda_0$ compared with the "0 slots" design. The coherence lengths here are sufficient to realize a certain level of wavefront control[54], and they can potentially be further improved by increasing the modal lifetime or engineering the band structure[69]. Note that a large coherence length also indicates high susceptibility to angular broadening, i.e., spectral broadening when emitted signals are collected from a range of angles, which we tried to avoid in our application (NDIR). For the "0 slots" and "1 slot" designs, the linewidth change caused by 10° incident angle averaging is 33 cm$^{-1}$ to 35 cm$^{-1}$ and 22 cm$^{-1}$ to 29 cm$^{-1}$, respectively. More detailed analysis can be found in Section S4 of the SI. In summary, our quasi-BIC designs could provide a well-balanced spatial coherence and angular broadening characteristics.

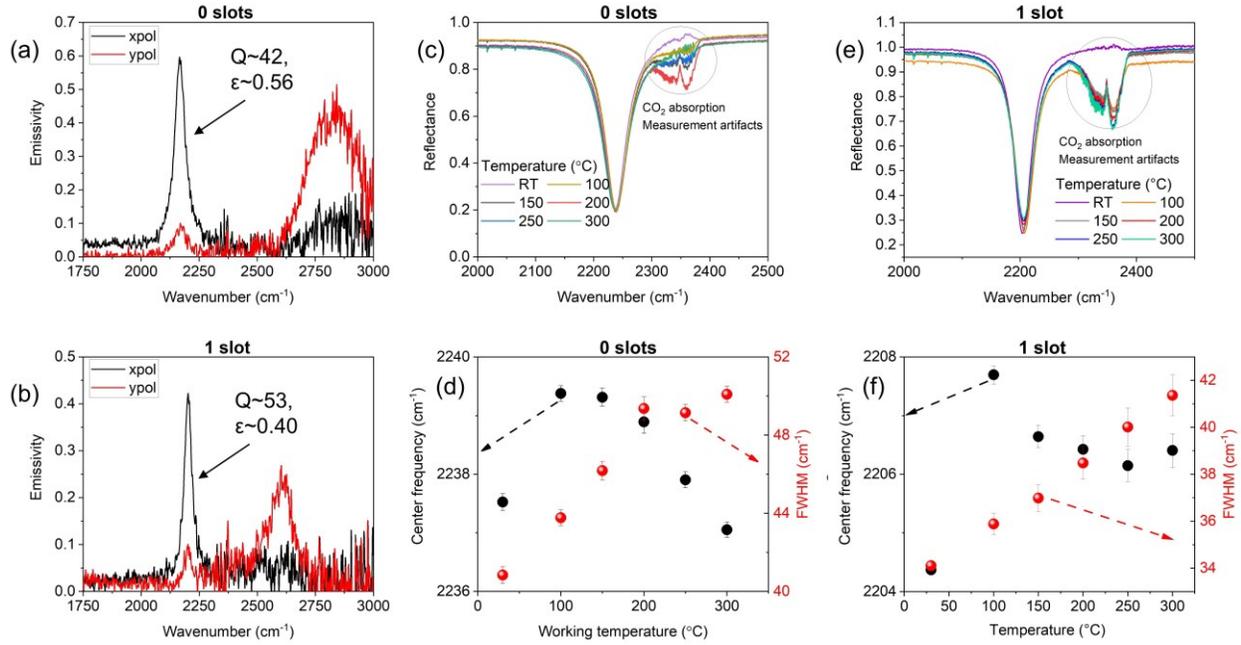

**Figure 4.** Spectroscopy measurements on fabricated samples. All measurements were conducted on patterns measuring 0.5 mm by 0.5 mm. (a) Measured emission at 300°C of $x$ and $y$ polarizations for the "0 slots" configuration. The primary quasi-BIC peak (black solid curve) shows a Q factor of ~ 42 and emissivity of ~ 0.56. (b) Measured emission at 300°C of $x$ and $y$ polarizations for the "1 slot" configuration. The primary quasi-BIC peak (black solid curve) shows a Q factor of ~ 53 and emissivity of ~ 0.40. Emissivity ($\varepsilon$) was measured by normalizing the emitted power from the quasi-BIC area to a blackbody sample with the same size. More details are given in the Methods section of the SI. (c) Measured reflection at varying temperatures of $x$ polarization for "0 slots" configuration. (d) Extracted center frequency and full width at half maximum (FWHM) from (c). (e) Measured reflection at varying temperatures of $x$ polarization for "1 slot" configuration. (f) Extracted center frequency and FWHM from (e).

To substantiate our design principle, we fabricated square patterns measuring 0.5 mm by 0.5 mm, with representative SEM images presented in Figures 1b and d. We then analyzed the emission spectra using a FTIR microscope (Bruker Corporation). Further details about fabrication and measurement procedures are elaborated in the Methods section of the SI. As depicted in Figures 4a and b, both the "0 slots" and "1 slot" configurations exhibit a single peak and narrow-band emission peak with $x$ polarization at 300°C. More specifically, the "0 slots" sample presents a Q factor of approximately 42 and a peak emissivity of 0.56, while the "1 slot" sample shows a Q factor of approximately 53 and a peak emissivity of 0.40. The *y*-polarized emission demonstrates a substantially broader bandwidth than the simulations, primarily due to the large angular dispersion of that mode. In order to confirm the temperature-stable resonance variation of the metallic design's optical response, we measured the spectra of the sample under varying temperatures. For a more accurate comparison with the performance of the sample at room temperature, we measured the reflection instead of the emission. As depicted in Figures 4c to f, the center frequency exhibits a minimal shift (~ 1 cm$^{-1}$), thus corroborating the temperature-independence of this design. However, the full width at half maximum (FWHM) of the spectra slightly broadens during the temperature increases from room temperature to 300°C, attributable to enhanced material losses.

The noticeable performance deterioration of the fabricated samples in comparison to the numerical simulations - for instance, reduced emission/absorption and a lower Q factor - can be attributed to several factors. Firstly, the finite pattern size and fabrication uncertainties (such as a slot width different from 100 nm, see Figures 5d and e) may impair the designed quasi-BIC resonance. The less-than-ideal quality of our deposited gold film could increase the material losses, leading to a reduction in the Q factor of the dissipative losses and a deviation from the CC condition. Moreover, the angular illumination and light collection introduced by our objective lens (with a numerical aperture NA = 0.17) generally tend to broaden the measured spectra. This expansion is primarily due to the non-flat band structure, a topic we have elaborated on in Figure 3. Note that the reflection and emission were conducted in two setups, i.e., different beam paths, and the beam divergences in reflection and emission measurements are different yet difficult to quantify. Therefore, the measured emissivity is not strictly equivalent to (1 - reflectivity). Performance enhancements can be achieved by addressing the aforementioned issues, such as fabricating millimeter-scale patterns and improving the gold deposition process. Overall, such a performance level is unattainable in conventional metallic designs. This is illustrated in Table 1, where we compare our design with several experimentally measured MIM-based EMs. To illustrate potential applications, we numerically demonstrate the highly sensitive detection of $N_2O$ gas by using our fabricated EM directly as a mid-infrared light source (see Section S5 of the SI). With the "0 slots" configuration as the benchmark, the sensitivity of the "1 slot" design is compromised (15%) because of the relatively small signal intensity (linewidth), but the selectivity is improved by ~ 30% since less signal can be absorbed by interference gases (CO here) for higher Q factors. If CO is not present in the target application, one could choose the "0 slots" design for an improved sensitivity, while the slot design is preferred when CO could coexist with the target gas. Notably, this process does not require any bandpass filters due to the inherent narrow-band property.

Table 1 Comparison between Different Experimentally Measured MIM-configured EMs.

| Structure | Wavelength (μm) | Q factor | Temperature (K) | Journal |
|---|---|---|---|---|
| Al MIM | 4.2 | 14 | 373 | ACS Photonics 2015[70] |
| Au MIM | 4.25 | ~ 6.2 | 873 | Phys. Rev. Appl. 2015[12] |
| Au MIM | 5.8 | < 10 | 573 | Phys. Rev. Lett. 2011[9] |
| Au MIM | ~ 4.0 | 11 | 573 | Appl. Phys. Lett. 2014[10] |
| Cu MIM | 3.96 | 15.7 | 603 | ACS Photonics 2017[4] |
| Cu MIM | ~ 4.26 | 15.8 | 673 | Nano Lett. 2020[5] |
| Au MIM (BIC) | 4.5 | 42/53 | 573 | This work |

Next, we illustrate how engineering the air slots can offer an additional degree of freedom for modulating the quasi-BIC resonance and hence, the emission properties in such a plasmonic system. As discussed earlier, in this plasmonic system, the induced ED moments are a product of circulating displacement currents on the metal resonators, unlike the induced electric displacement field in the dielectric system[63]. As evident from Figure 5c, when the slot is not long enough to cut-through the gold elliptical resonator, the spectra show only minor alterations. This is because the circulating currents are able to navigate around the slot's edges when it is not fully cut through. Only when the slot completely bisects the resonator, interrupting the original circulating currents, does a significant resonance peak shift towards larger frequencies become noticeable, along with a substantial increase in the Q factor, thus yielding a narrower linewidth. The drop in absorptance ($A = 1 - R$) is primarily attributed to a deviation from the CC condition.

As reported in our previous work[63], the spectral responses of the slotted design can be interpreted through the optical nanoantenna theory[71–75]. Similarly, we employ an effective impedance model here to comprehend and forecast the spectral responses of the slotted design. For simplicity, we restrict our attention to a single gold elliptical resonator in one unit cell. It has been demonstrated that the near-field electric polarization's dominant contribution stems from the electric dipole moment along the ellipse's long axis. Assuming the resonator only experiences radiative losses, a single dipolar nanoantenna can generally be approximated with a series RLC circuit (refer to Figure 5a). This is associated with the system's intrinsic impedance, expressed as

$$Z_{int} = R_{int} - iX_{int} = R_{int} - i\omega L_{int} - \frac{1}{i\omega C_{int}} \qquad (1)$$

where i represents the imaginary unit, and $R_{int}$ refers to the intrinsic resistance, which is due to radiative losses. According to Alù et al[72], a dipolar nanoantenna with a centrally loaded gap can function as a series combination of the intrinsic impedance and the load impedance when excited by an external plane wave, provided that the electric field runs parallel to the nanodipole's axis. Consequently, the input impedance of the loaded nanoantenna can generally be expressed as

$$Z_{in} = R_{in} - iX_{in} = Z_{int} + Z_{slot} \qquad (2)$$

where in our case the load impedance can be expressed as

$$Z_{\text{slot}} = \frac{iw}{\omega \text{Re}[\varepsilon] H b} \quad (3)$$

In this context, *H*, *w* and *b* represent geometric parameters, as depicted in Figure 1. The optical impedance is capacitive for dielectric gap loading materials that have positive values for the real part of the permittivity. However, for metals, the real part of the permittivity can assume a negative value within specific frequency bands. For instance, noble metals such as silver and gold exhibit plasma frequencies in the visible or ultraviolet range. Consequently, they have a negative real part of the permittivity in optical and infrared frequencies, thereby demonstrating plasmonic properties. In such cases, the equivalent impedance of the metal nanoloads can appear as "negatively capacitive" for any frequency where $\text{Re}[\varepsilon] < 0$ and can be perceived as a positive effective inductance[71,72]. For this series RLC circuit, the Q factor is defined as follows

$$Q = R_{\text{int}}^{-1} \sqrt{L_{\text{int}} / C_{\text{system}}} \propto \sqrt{1/C_{\text{system}}} \quad (4)$$

Here, $C_{\text{system}}$ is the series connection of $C_{\text{int}}$ and $C_{\text{slot}}$ expressed as:

$$C_{\text{slot}} = \text{Re}[\varepsilon] H b / w \quad (5)$$

$$C_{\text{system}} = (1/C_{\text{int}} + 1/C_{\text{slot}})^{-1} = \frac{1}{\left(\frac{1}{C_{\text{int}}} + \frac{w}{\text{Re}[\varepsilon] H b}\right)} \quad (6)$$

Now we can decipher the radiative Q factor responses of the slotted design. When the slot is filled by dielectric materials, the increase in permittivity leads to an increased $C_{\text{system}}$, consequently reducing the Q factor. This behavior is distinctly seen in Figure 5b, where the air slot (i.e., the "1 slot" configuration) exhibits the narrowest linewidth and highest Q factor amongst all variations of permittivity for the material filling the slot. As the Q factor decreases, the system reverts to the CC condition, which explains the near-unity absorptance at higher permittivities. Conversely, when the filling material is switched to gold, which is the case for the "0 slots" configuration, the negative real part of the gold's permittivity causes $C_{\text{system}}$ to increase even further. This is substantiated by the dashed curve in Figure 5b, which corresponds to a gold-filled slot and exhibits the lowest Q factor. For slots filled with dielectric materials, as reported in our previous work[63], the resonance peak correlates with the open-circuit resonance frequency (when $R_{in}$ is a maximum and $X_{in} = 0$) expressed as[63,74]

$$\omega_0 = \frac{1}{X_{int}C_{slot}} \quad (7)$$

Therefore, the increased permittivity leads to an increased $C_{slot}$ (as per Eq. (5)), thereby shifting the resonance peak to lower frequencies, as depicted in Figure 5b. For the "0 slots" configuration (i.e., the slot is filled with gold), since the resonance shifts to near the short-circuit resonance when plasmonic materials are employed to fill the gap[72], Eq. (7) cannot be employed to interpret the spectrum shift. However, the short-circuit resonance correlates with significantly lower frequencies, which elucidates the substantial redshift (i.e., shifting towards a smaller wavenumber) seen in Figure 5b for the dashed curve.

By modulating the slot width (*w*), we can also manipulate the critical coupling between radiative and dissipative losses, thereby fine-tuning the Q factor and absorption/emission characteristics. This is numerically represented in Figure 5d and corroborated experimentally in Figure 5e. As we observe, a larger width (*w*) shifts the peak towards higher frequencies. The radiative Q factor increases, resulting in a higher total Q factor (i.e., $Q_{tot} \sim 77$ for *w* = 50 nm and $Q_{tot} \sim 166$ for *w* =

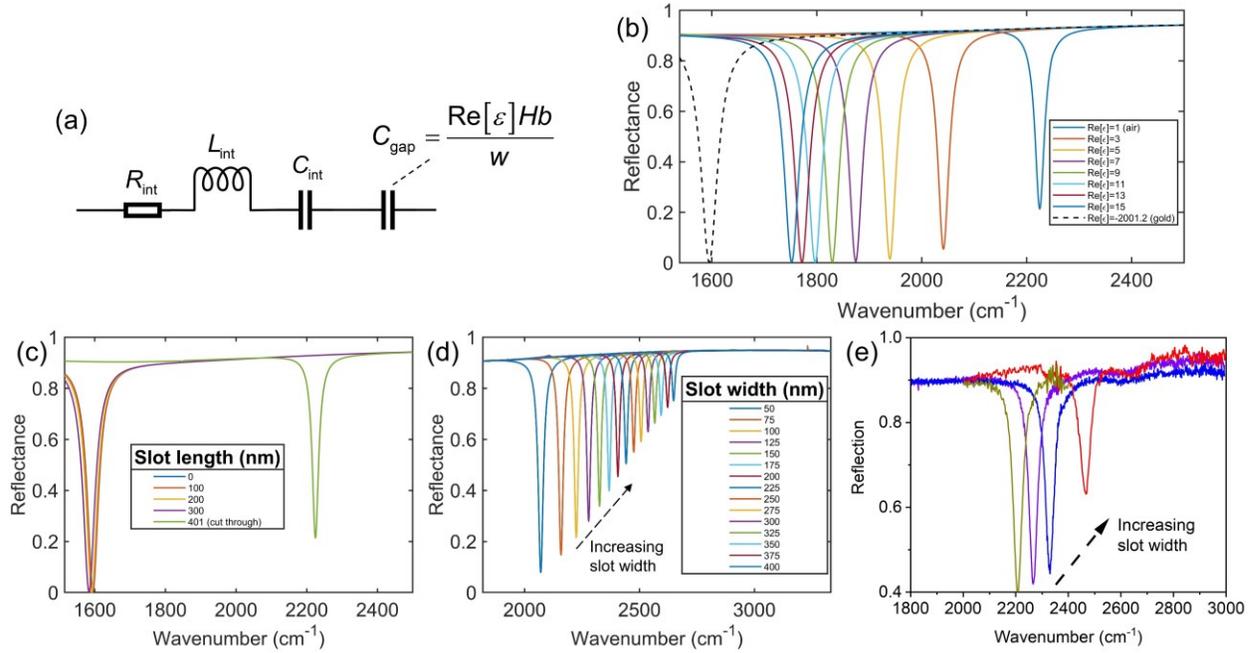

Figure 5. Impacts of slot properties and effective impedance model. (a) Equivalent circuit model for the "1 slot" configuration. (b) Numerically calculated spectra with respect to different permitivities of the slot-filling material. As an example, the "1 slot" configuration refers to filling the slot with a material having a real part of permittivity (Re[$\varepsilon$]) of 1, while the "0 slots" configuration corresponds to filling the slot with gold. Geometrical parameters of the slot are: $w$ = 100 nm and $l$ = 401 nm (cut through the gold ellipse). (c) Numerically calculated spectra for varying slot lengths ($l$) with a fixed slot width $w$ = 100 nm. The solid lines correspond to lengths from $l$ = 0 nm ("0 slots" configuration) to $l$ = 401 nm, left to right. (d) Numerically calculated spectra for different slot widths ($w$) when the slot cuts through the gold ellipse ($l$ = 401 nm). The solid lines represent widths from $w$ = 50 nm to $w$ = 400 nm, left to right. (e) Experimentally measured spectra for various slot widths ($w$) with the slot cutting through the gold ellipse. Measurements were conducted on patterns measuring 200 μm by 200 μm at 300°C.

400 nm), and subsequently, lower absorptance due to a deviation from the CC condition. This behavior can also be well forecasted using the effective impedance model. As demonstrated by Eq. (5)-(6), an increased $w$ leads to both a reduced $C_{\text{slot}}$ and $C_{\text{system}}$. Consequently, we observe resonances with a higher Q factor and a shift of the peak to higher frequencies, as predicted by Eq. (4) and (7), respectively. It is worth noting that the influence of the slot width is more pronounced in this system than in lossless dielectric systems[63], owing to the critical coupling between radiative and dissipative losses. In this context, we have successfully extended our slot engineering methodology to dipole-based plasmonic resonant systems.

## 3. Conclusion

In conclusion, we have successfully demonstrated, for the first time, a mid-infrared, wavelength-selective thermal emitter driven by plasmonic quasi-BIC modes. Our design showcases a measured narrow emission bandwidth and a recorded Q factor of approximately 42, a level of performance that traditional metal-insulator-metal-based thermal emitters fail to achieve. Furthermore, the plasmonic quasi-BIC mode dominates over a wide wavelength range, enabling a singular emission peak within a broad spectral range of at least 10 μm (3 μm to 13 μm). The peak position is adjustable and can be precisely controlled by tailoring the quasi-BIC mode. By introducing symmetric air slots into the metallic resonators, the Q factor can be enhanced by about 22%, and further tunability can be attained through careful engineering of the slot. Spectral measurements substantiate the resilience of our designs' resonance frequency to temperature fluctuations, opening new avenues for the use of plasmonic quasi-BICs in designing ultra-narrowband and temperature-stable thermal emitters. Replacing gold with refractory plasmonic materials allows for high-temperature operation. Moreover, fine-tuning the BIC modes can yield other controls including polarization, angular dispersion, and far-field profile[53], establishing this platform as an ideal mid-infrared light source. Additionally, this concept can be extended to other frequencies for a range of applications, such as near-infrared for thermophotovoltaic energy conversion[50], Terahertz[76] for remote sensing of chemical and biological species, and Gigahertz for wire communications[77].

ASSOCIATED CONTENT

**Supporting Information**

AUTHOR INFORMATION

**Corresponding Author**

* **Justus C. Ndukaife** - Department of Electrical and Computer Engineering, Vanderbilt University, Nashville, Tennessee 37235, United States; Interdisciplinary Materials Science, Vanderbilt University, Nashville, Tennessee 37235, United States; Vanderbilt Institute of Nanoscale Science and Engineering, Vanderbilt University, Nashville, Tennessee 37235, United States; ORCID https://orcid.org/0000-0002-8524-0657; Email: justus.ndukaife@vanderbilt.edu

**Authors**

**Sen Yang** - Department of Electrical and Computer Engineering, Vanderbilt University, Nashville, Tennessee 37235, United States; Vanderbilt Institute of Nanoscale Science and Engineering, Vanderbilt University, Nashville, Tennessee37235, United States; ORCID https://orcid.org/0000-0002-0056-3052

**Mingze He** - Department of Mechanical Engineering, Vanderbilt University, Nashville, Tennessee 37235, United States; Vanderbilt Institute of Nanoscale Science and Engineering, Vanderbilt University, Nashville, Tennessee 37235, United States; ORCID https://orcid.org/0000-0001-8773-1268

**Chuchuan Hong** - Department of Electrical and Computer Engineering, Vanderbilt University, Nashville, Tennessee37235, United States; Vanderbilt Institute of Nanoscale Science and Engineering, Vanderbilt University, Nashville, Tennessee37235, United States; ORCID https://orcid.org/0000-0002-1329-9385

**Josh Nordlander** - Department of Materials Science and Engineering, The Pennsylvania State University, University Park, Pennsylvania 16802, United States

**Jon-Paul Maria** - Department of Materials Science and Engineering, The Pennsylvania State University, University Park, Pennsylvania 16802, United States

**Joshua D. Caldwell** - Department of Mechanical Engineering, Vanderbilt University, Nashville, Tennessee 37235, United States; Interdisciplinary Materials Science, Vanderbilt University, Nashville, Tennessee 37235, United States; Vanderbilt Institute of Nanoscale Science and Engineering, Vanderbilt University, Nashville, Tennessee 37235, United States; ORCID https://orcid.org/0000-0003-0374-2168

**Author Contributions**

J.C.N. guided the project. S.Y. and M.H. conceived the idea. S.Y. designed the plasmonic quasi-BIC thermal emitter. S.Y. performed the electromagnetic simulation, and fabricated the metasurface. M.H. designed and conducted the spectral measurement. C.H. conducted the multipole decomposition analysis and contributed to the metasurface fabrication. J.N. prepared the substrate for holding the metasurface. J.-P.M. guided the substrate fabrication. J.D.C. guided the spectral measurement experiment. All authors contributed to the manuscript editing.

**Author Contributions**

S.Y. and M.H. contributed equally to this work.

**Notes**

The authors declare no competing financial interest.

ACKNOWLEDGMENT

S.Y., C.H., and J.C.N. acknowledge financial support from the National Science Foundation NSF CAREER Award (NSF ECCS 2143836). J.D.C. acknowledges support from NSF #2128240, while M.H. was supported by the Office of Naval Research under grant #N00014-22-1-2035. J.N. gratefully acknowledges support from the Department of Defense (DoD) through the National Defense Science and Engineering Graduate (NDSEG) Fellowship Program. J.-P.M. Fabrication of the quasi-BIC metasurfaces was conducted at the Vanderbilt Institute of Nanoscale Science and Engineering.

# Supporting Information

# Single-peak and narrow-band mid-infrared thermal emitters driven by mirror-coupled plasmonic quasi-BIC metasurfaces

*Sen Yang[1,3,‡], Mingze He[2,‡], Chuchuan Hong[1], Josh Nordlander[4], Jon-Paul Maria[4], Joshua D. Caldwell[2,3,5], and Justus C. Ndukaife[1,3]\**

[1]Department of Electrical and Computer Engineering, Vanderbilt University, Nashville, TN, USA 37235

[2]Department of Mechanical Engineering, Vanderbilt University, Nashville, TN, USA 37235

[3]Interdisciplinary Materials Science Program, Vanderbilt University, Nashville, TN, USA 37240

[4]Department of Materials Science and Engineering, The Pennsylvania State University, University Park, Pennsylvania, USA 16802

[5]Sensorium Technological Labs, 6714 Duquaine Ct, Nashville, TN, USA 37205

‡These authors contribute equally to this work.

*Correspondence: justus.ndukaife@vanderbilt.edu

**Manuscript title:** Single-peak and narrow-band mid-infrared thermal emitters driven by mirror-coupled plasmonic quasi-BIC metasurfaces

**Pages:** S1 – S9

**Figures:** S1 – S7

**Sections:** S1 – S6

# S1 Multipolar characters of the quasi-BIC mode

We employed multipolar decomposition projected onto Cartesian coordinates as an analytical tool to examine the far-field radiation of the plasmonic quasi-BIC mdoes[1-4]. Our first step was to calculate the induced current density as a function of the near-field distribution adjacent to the resonator. The near-field distributions were derived using a commercial electromagnetic solver (Lumerical FDTD). We should note that there is a minor discrepancy (several cm$^{-1}$s) between the calculated resonance frequencies and those obtained from the spectra produced by CST Studio.

The induced current density, $\mathbf{J}(\mathbf{r})$, is expressed as follows:

$$\mathbf{J}(\mathbf{r}) = i\omega\varepsilon_0(n^2 - 1)\mathbf{E}(\mathbf{r}) \qquad (1)$$

, where $\omega$ is the angular frequency, $\varepsilon_0$ is the permittivity of free space, and $n$ is the refractive index of the resonator. $\mathbf{J}(\mathbf{r})$ is associated with the displacement current distributions. The first four orders of the multipolar components, including electric dipole ($p_\alpha$), magnetic dipole ($m_\alpha$), electric quadrupole ($\hat{Q}^e_{\alpha\beta}$), and magnetic quadrupole ($\hat{Q}^m_{\alpha\beta}$), are expressed by:

$$p_\alpha = -\frac{1}{i\omega}[\int J_\alpha j_0(kr)\, d^3\mathbf{r} + \frac{k^2}{2}\{3(\mathbf{r}\cdot\mathbf{J})r_\alpha - r^2 J_\alpha\}\frac{j_2(kr)}{(kr)^2}d^3\mathbf{r}] \qquad (2)$$

$$m_\alpha = \frac{3}{2}\int(\mathbf{r}\times\mathbf{J})_\alpha \frac{j_1(kr)}{kr}d^3\mathbf{r} \qquad (3)$$

$$\hat{Q}^e_{\alpha\beta} = -\frac{3}{i\omega}[\int\{3(r_\beta J_\alpha + r_\alpha J_\beta) - 2(\mathbf{r}\cdot\mathbf{J})\delta_{\alpha\beta}\}\frac{j_1(kr)}{kr}d^3\mathbf{r} + 2k^2\int\{5r_\alpha r_\beta(\mathbf{r}\cdot\mathbf{J}) - r^2(r_\beta J_\alpha + r_\alpha J_\beta) - r^2(\mathbf{r}\cdot\mathbf{J})\delta_{\alpha\beta}\}\frac{j_3(kr)}{(kr)^3}] \qquad (4)$$

$$\hat{Q}^m_{\alpha\beta} = 15\int\{r_\alpha(\mathbf{r}\times\mathbf{J})_\beta + r_\beta(\mathbf{r}\times\mathbf{J})_\alpha\}\frac{j_2(kr)}{(kr)^2}d^3\mathbf{r} \qquad (5)$$

where $\alpha, \beta = x, y, z$ and $k$ is the wavenumber. $j_n(\rho)$ denotes the spherical Bessel function defined by $j_n(\rho) = \sqrt{\pi/2\rho}\, J_{n+1/2}(\rho)$, where $J_n(\rho)$ is the Bessel function of first kind.

This multipolar analysis contains the electric field information contributed by the gold elliptical resonators, the DyF$_3$ spacer layer and the CdO reflector layer, as depicted in Figure S1, where two different cases are investigated. The decomposed results are presented in Figures S2.

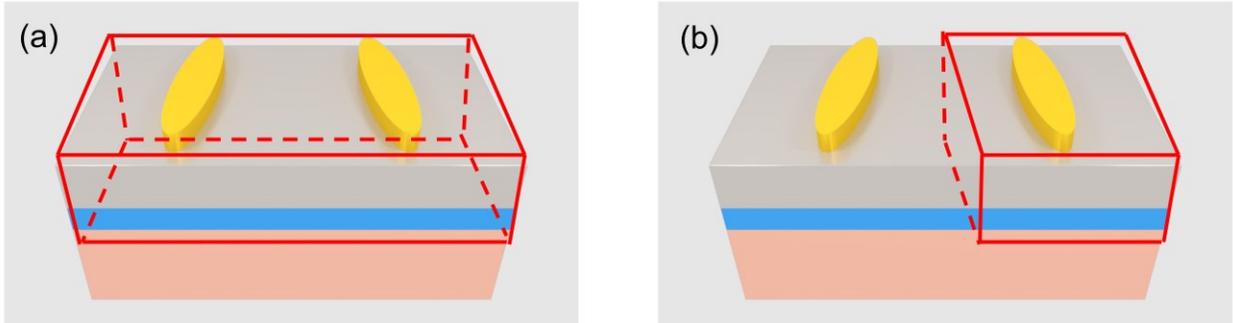

Figure S1. Schematic illustration highlighting the region within a unit cell, as demarcated by the red dashed rectangle, to which multipolar decomposition analysis was applied for (a) a pair of resonators and (b) a single resonator.

To gain deeper insight into the multipolar characteristics of the quasi-BIC mode before and after the introduction of air slots, we perform an integration of the electric fields over half a unit cell (which contains a single gold elliptical resonator, as depicted in Figure S1b) and the entire unit cell (which houses two gold elliptical resonators, as shown in Figure S1a), respectively. For the "0 slots" configuration, the dominant $y$ component of ED in a single ellipse (as shown in Figure S2a), the predominant $z$ component of MD and the strong EQ exhibited for the whole unit cell (as presented in Figure S2b), all present the characteristic features common to similar tilted-ellipse-based all-dielectric quasi-BIC metasurfaces[4]. Upon introducing the air slots, there is a considerable increase in the $x$ component of the MD for half a unit cell, and a notably strong in-plane MQ appears in one unit cell, attributed to the evolution from one to two parallel magnetic dipole moments beneath each gold elliptical resonator. The radiation pattern thus evolves from a dipole mode to a high-order mode, featuring the higher Q factor[5].

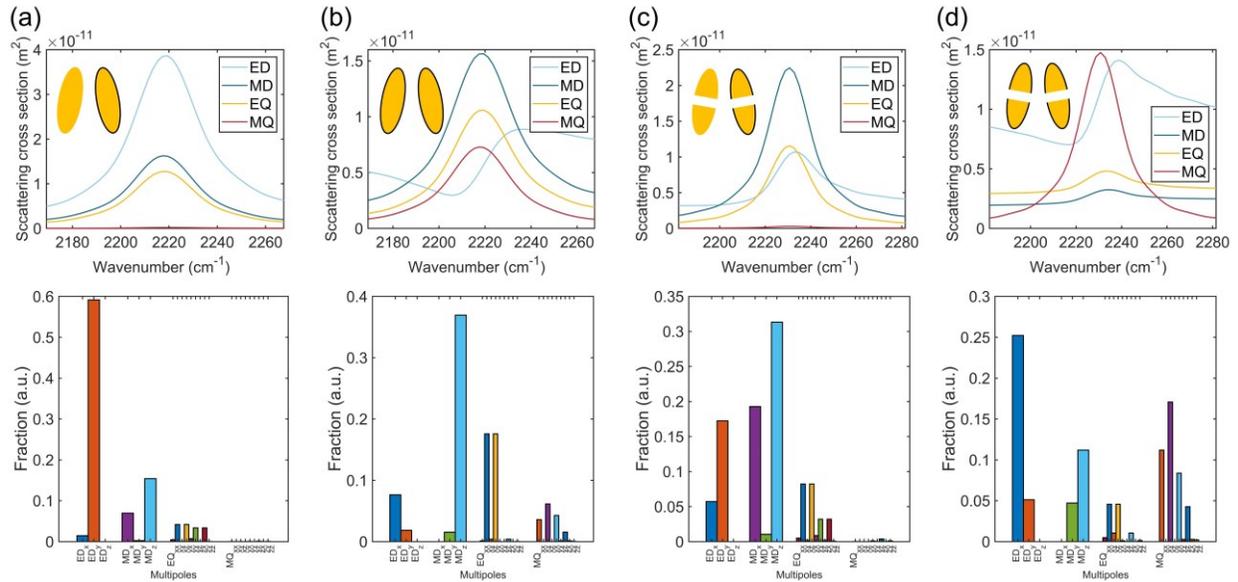

Figure S2. Multipolar decomposition of the quasi-BIC resonance. The first row shows the contributions from the electric dipole (ED), magnetic dipole (MD), electric quadrupole (EQ) and magnetic quadrupole (MQ) components in the chosen region (marked by black outlines and corresponding to Figures S1a or b) as a function of wavenumbers. The second row shows the corresponding Cartesian fractions of all multipolar contributions for the chosen region at quasi-BIC resonance frequency. (a) and (b) are for the "0 slots" configuration. (c) and (d) are for the "1 slot" configuration.

For a more comprehensive understanding of the multipolar decomposition results, we delve into the examination of electromagnetic field distributions for the two distinct configurations. As depicted in Figure S3a, the electric field is markedly enhanced and localized around the tips of the gold elliptical resonators, with a peak enhancement factor of approximately 58. The introduction of air slots further escalates the electric field, achieving a maximum enhancement factor of roughly 75. This introduces additional regions of electromagnetic enhancement and confinement within the slot, as shown in Figure S3e. Comparing Figure S3b with S3f, as well as Figure S3c with Figure S3g, we clearly corroborate the conclusion mentioned in the main manuscript that the circulating current induced on each tilted gold elliptical resonator is disrupted and split into two separate streams, leading to the evolution from one to two parallel magnetic dipole moments induced in the spacer layer beneath each resonator, as shown in Figures S3d and S3h. The magnetic dipole moments beneath the left and right elliptical resonators within the unit cell also display the characteristic anti-parallel behavior in these tilted-ellipse-based, symmetry-protected quasi-BIC systems[6–8].

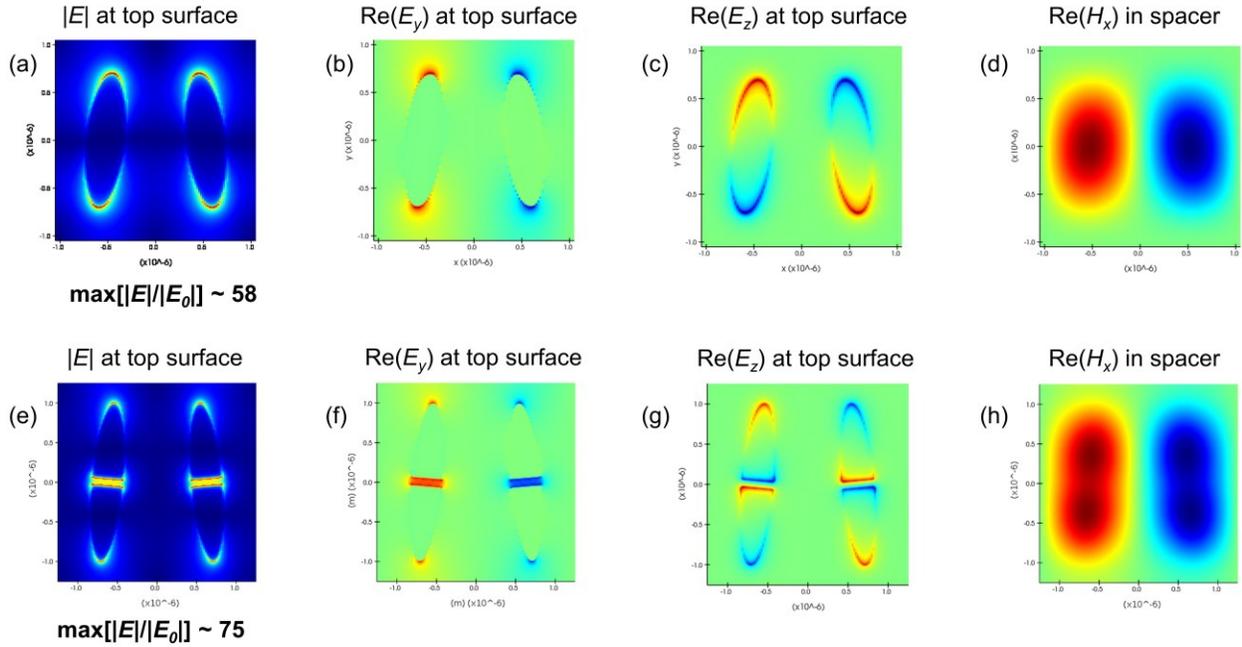

Figure S3. Electromagnetic field distributions (*xy* plane) for "0 slots" and "1 slot" configurations on quasi-BIC resonance. (a) Electric field enhancement distribution of the *xy* plane ($z = H$) for the "0 slots" configuration in one unit cell. The maximum field enhancement is 58, which is defined as the maximum electric field enhancement factor on the *xy* plane. (b) Distributions of the *y* component of the electric field, (c) *z* component of the electric field and (d) *x* component of the magnetic field for the "0 slots" configuration in one unit cell. (d) Electric field enhancement distribution of the *xy* plane ($z = H$) for the "1 slot" configuration in one unit cell. The maximum field enhancement is elevated to 75. (f) Distributions of the *y* component of the electric field, (g) *z* component of the electric field and (h) *x* component of the magnetic field for the "1 slot" configuration in one unit cell.

## S2 Optical properties of the CdO film

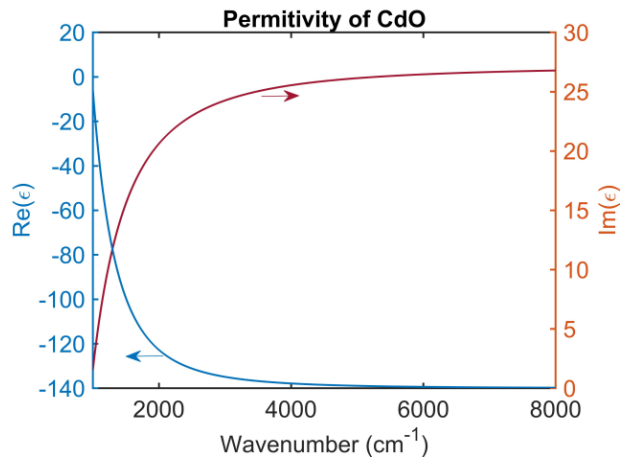

Figure S4. Measured permittivity of CdO at room temperature. The same data was used in all electromagnetic simulations.

## S3 Angular dispersion of the thermal emission

The quasi-BIC mode of the metasurface can be excited provided there is a nonzero electric field component along the *x*-axis from the incident light. Consequently, we can alter the light incident angle in two distinct planes of the unit cell (*xz* and *yz*) to excite the quasi-BIC mode[8]. For ease of understanding, we have termed these two approaches as TM mode and TE mode, respectively, as depicted in Figure S5a. The TM*x* mode implies the wave vector and polarization in the *xz* plane, whereas the TE*y* mode describes light with a wave vector in the *yz* plane and polarization along the *x*-axis. Given that the metasurface displays mirror symmetry along the *y*-axis but not the *x*-axis, the optical responses affiliated with the quasi-BIC mode fluctuate between the TM and TE modes. This means that the thermal emitter's emission properties differ between TM and TE modes. Figures S5b to 5e depict the angular dispersion properties of the TM and TE modes for the "0 slots" and "1 slot" configurations, respectively. Two main conclusions can be drawn from these figures, further reinforced by Figure S5f: (1) The "0 slots" configuration presents a flatter band in comparison to the "1 slot" configuration, implying that the thermal emission from the "1 slot" configuration possesses superior directivity. (2) The TE mode features a flatter band compared to the TM mode, leading to a smaller angular dispersion. It is noteworthy that the split lobes at higher incident angles for the TE mode of the "1 slot" configuration (as per Figure S5e) are a result of the interaction between the original quasi-BIC mode and a newly emerged mode.

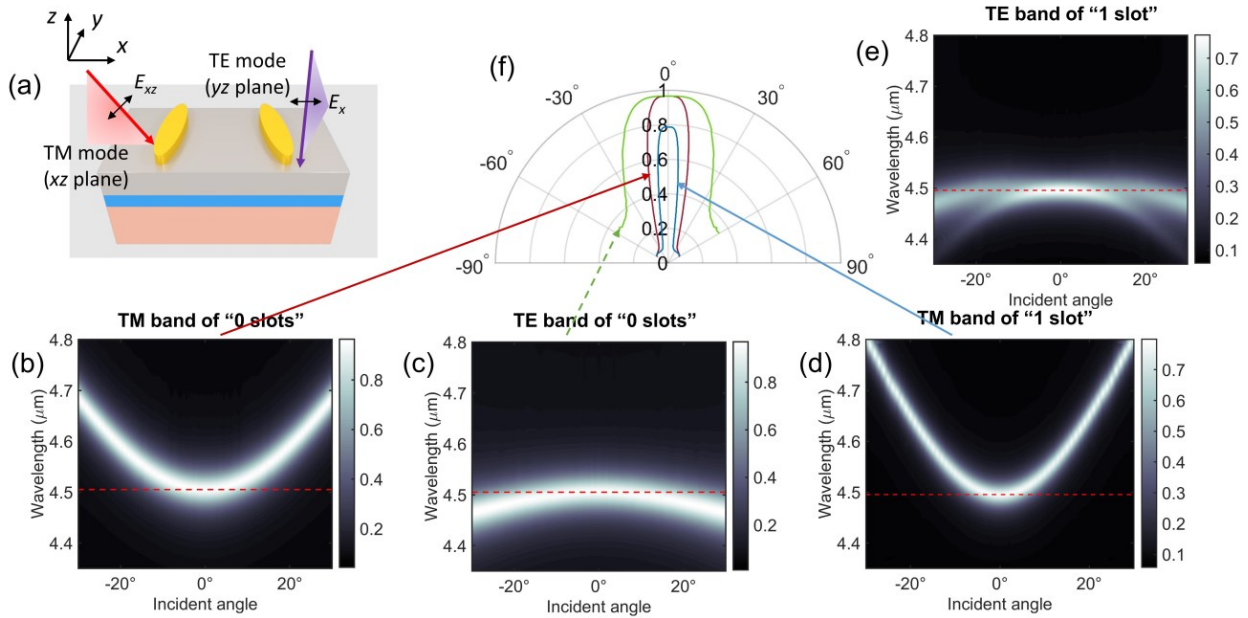

Figure S5. Angular dispersion of TE and TM modes for "0 slots" and "1 slot" configurations. (a) Schematic illustration showing how the TE and TM modes are defined. (b) Band diagrams of the "0 slots" configuration showing the absorptance spectra as a function of incident angles for TM mode and (c) for the TE mode. (d) Band diagrams of the "1 slot" configuration showing the absorptance spectra as a function of incident angles for TM mode and (e) for the TE mode. (f) Polar diagram comparing the angular dispersion properties of absorption (i.e., emission) for the "0 slots - TM", "0 slots - TE" and "1 slot - TM" configurations.

## S4 Spectral broadening in measurements due to angular dispersion

While the dispersion leads to spatial coherence that are beneficial to wavefront engineering[9], e.g., lensing[10] and polarization control[11], it complicates our measurements. In the actual measurement, including reflection and emission, the response of the sample was collected in a spanning of angles instead of a singular angle. Since we used a near-normal refractive objective to collect the signal, the range of the incident angle is approximately spread from 0 to 10°, and the beam is collected and passes through an iris to further reduce the divergence. As we do not know the accurate angular distribution, here we estimated it via two approaches: (1) assuming that light from 0-10° evenly contributes to the signal; (2) assuming that the signal is a gaussian distribution with a bandwidth of 5°, as shown in Figure S6c. Then based on the contribution of signal at different incident angles, we can estimate the spectral broadening, as shown in Figures S6a and b for the "0 slots" and "1 slot" configurations, respectively. We find that the "0 slots" configuration is generally not susceptible to angular divergence within 0-10°, while the performance of "1 slot" configuration is influenced significantly due to the sharper band curvature (see Figure S5). Therefore, the experimentally measured full width at half maximum (FWHM) of the "1 slot" design could be reduced more significantly than the "0 slots" design.

.

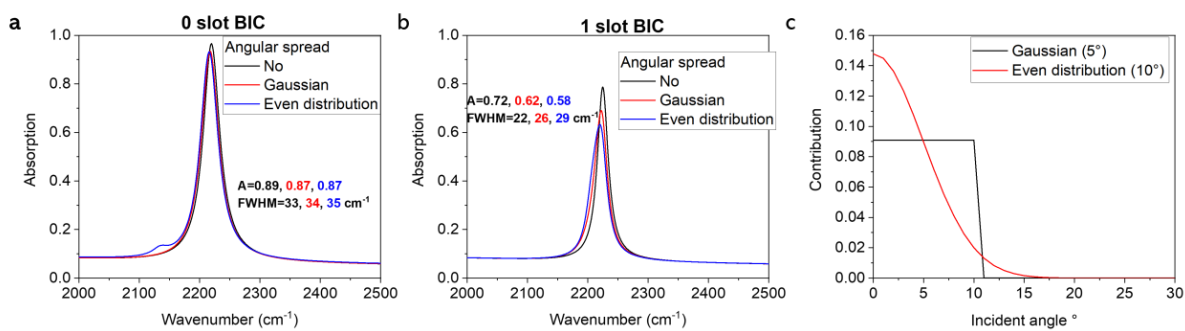

Figure S6. Estimation of the angular dispersion in measurements. (a) Numerically calculated absorption of the "0 slots" configuration, when considering "no angular divergence", "gaussian distributed divergence", and "evenly distributed divergence", respectively. (b) Same calculations for the "1 slot" configuration. (c) Contributions at different incident angles for the two estimation approaches.

## S5 Numerical analysis on non-dispersive infrared (NDIR) gas detection

The working principle of filterless NDIR has been discussed in multiple references[12,13], and here we briefly discuss the fundamentals. In a filterless NDIR configuration, the emitted light from the light source passes through the gas cell, where the gases circulate through. The emitted energy would be absorbed by the gas based on the concentration and optical path length. Here, the numerical evaluations for the filterless NDIR were performed in two aspects: (1) sensitivity, i.e., absolute power change induced by a given concentration of target gas; (2) selectivity, i.e., the ratio of power change between the target gas ($N_2O$ here) and interference gas (CO in this case). To generalize the comparison, we shifted the center of the spectra for both the "0 slots" and "1 slot" thermal emitters to be at the same frequency. *Note that the data used in these calculations are experimentally measured data with manually shifted resonances.*

We firstly determine the optimal working frequency of the two configurations by calculating their corresponding sensitivities. We find that the optimal working frequency of the quasi-BIC for the two configurations should both be located at the second peak (2235 cm$^{-1}$) of the N$_2$O absorption spectrum, as shown in Figure S7a. In the numerical simulations, we manually shifted the working frequency of both configurations to be at 2235 cm$^{-1}$ and evaluated the sensitivities and selectivities. This can be achieved in practice by carefully adjusting the geometries of the resonators. Since the spectrum of the "1 slot" design has a smaller overlap with the gas absorption spectrum due to a narrower FWHM and a lower absorptance (Figure S7c), it leads to a decreased sensitivity compared with the "0 slots" design (Figure S7b red versus black solid lines). However, such a spectral feature also indicates a significantly reduced overlap with the CO absorption spectrum (Figure S7c), improving the selectivity by ~30% (Figure S7b red versus black dash lines). In actual product designs, it is eventually a trade-off, i.e., whether sensitivity or selectivity is prioritized for the specific application. Our design strategies thus provide a versatile toolkit to adjust the quality factors of quasi-BIC thermal emitters without significantly compromising the emissivity.

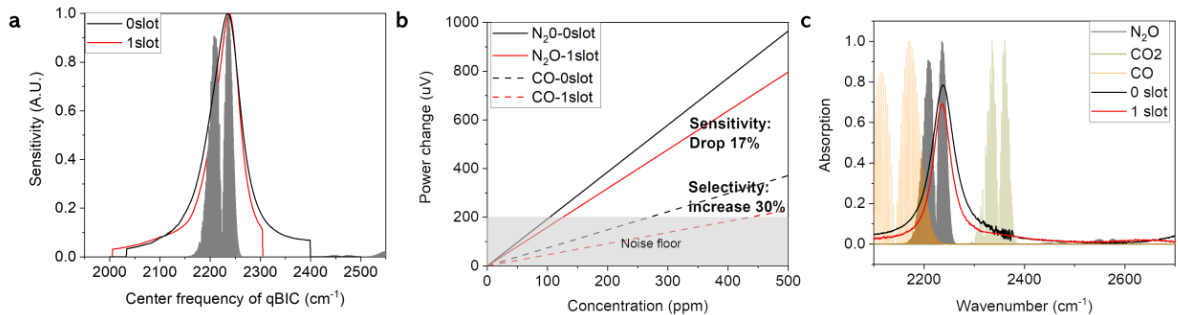

Figure S7. Evaluation of filterless NDIR gas detection performance. (a) Sensitivities of the two configurations ("0 slots" and "1 slot") to N$_2$O with respective to different center frequencies. The absorption spectrum of N$_2$O is plotted as the shaded region. (b) Power change with N$_2$O (target gas) and CO (interference gas which might lead to false positive). (c) Spectra overlaps for different gases and the two quasi-BIC thermal emitters. Gas spectra and absorption intensities are taken from HITRAN[13].

## S6 Methods

### S6.1 Simulations

Numerical simulations of the metasurface's optical responses were performed using the frequency domain finite element (FEM) Maxwell solver contained in CST STUDIO SUITE 2018 with a tetrahedral mesh. The field distributions are extracted from three-dimensional finite-difference time-domain (FDTD) simulations using a commercial software package (Ansys Lumerical FDTD Solutions 2021). We compared the simulation results obtained from both software to guarantee the differences between them are small enough.

The permittivity values of the DyF$_3$ spacer and sapphire (Al$_2$O$_3$) substrate were set as 2.83 (measured at 400°C) and 2.723, respectively. The permittivity value of gold was taken from Olmon et al.'s data[77]. The permittivity value of the CdO film was measured by ellipsometry, see Section S2 of the SI. The whole system was considered in air.

For simulations performed in CST STUDIO, the *unit cell* boundary conditions were set for *x* and *y* directions and the *open* boundary condition was set for the *z* direction. The *number of Floquet modes* was set as 18 to avoid fake peaks. The *Maximum step width* was 0.1 μm for the mirror, 0.05 μm for resonators and 0.025 μm for slots.

For simulations performed in Lumerical FDTD, *periodic* boundary conditions were set for *x* and *y* directions and the *PML* boundary condition was set for the *z* direction. The *Anti-Symmetric* boundary conditions were applied in the *x* direction to reduce computational cost. *Override mesh* regions enclosed the resonators, spacer and reflector layers with a *maximum mesh step* of 10 nm for *x*, *y* and *z* directions.

## S6.2 Fabrications

Sapphire-CdO-DyF3 substrate: In-doped CdO (n-type) was deposited on 2 in. r-plane (012) sapphire single crystal substrates at 400°C by a reactive cosputtering process employing high-power impulse magnetron sputtering (HiPIMS) and radio frequency sputtering from 2 in. diameter metal cadmium and indium targets, respectively. HiPIMS drive conditions were 800 Hz frequency and 80 μs pulse time, yielding a 1250 μs period and 6.4% duty cycle. Film growth occurred in a mixed argon (20 sccm) and oxygen (14.4 sccm) environment at a total pressure of 10 mTorr. Postdeposition, samples were annealed in a static oxygen atmosphere at 635°C for 30 min. Next, DyF$_3$ was deposited at ambient temperature using electron beam evaporation. Thickness was monitored throughout the deposition using a quartz crystal microbalance.

Plasmonic metasurface: The metasurfaces were fabricated on top of the prepared substrate. Spin coating of a double layer of polymethyl methacrylate (PMMA) resist of different molecular weights (495K A2 and 950K A4) was followed by resistive physical vapor deposition (Angstrom Amod) of a 10 nm chromium conduction layer. The 500 μm by 500 μm patterns were transferred to the resist by 30 keV electron beam lithography (Raith eLiNE), after which the sample was dipped first in the chromium etchant (Transene Chrome Mask Etchant 9030) to remove the conduction layer and then in the MIBK/IPA 1:3 (Kayaku Advanced Materials) for development. Afterwards, an adhesion layer of 3 nm titanium and 200 nm of gold were vertically deposited via electron-beam physical vapor deposition (Angstrom Amod) followed by the wet-chemical lift-off in NMP 1165 remover (Microposit) on an 80°C hot plate overnight.

## S6.3 Measurements

The samples were measured with the MCT detector of a Bruker VERTEX 70v FTIR spectrometer and a Hyperion 2000 microscope. A glass 5× (0.17 NA) objective was used for alignment of the devices, and a Ge refractive 5× (0.17 NA) objective was used for the reflectance and emission measurements. The reflectance was normalized to a gold mirror.

In the case of reflection measurements, a Globar built in the FTIR was used to illuminate the sample, and a motorized aperture was used to control the area of signal collected.

For thermal emission measurements, we turned off the Globar and directly collected the thermally emitted signal from the quasi-BIC devices, and the spectra were measured with FTIR, notated as $M_{sample}$. The emitted power from a blackbody (vertically aligned carbon nanotubes) at the same temperature was also measured ($M_{bb}$). An aperture was placed in the microscope so that only emission from the size of the array was collected in both measurements. Finally, we blocked the signal from the microscope and measured the emission spectrum from the instrument itself, noted as $M_{background}$. From there, the emissivity of the sample can be determined by the following equation:

$$\varepsilon(T_{\text{sample}}) = \frac{M_{\text{sample}} - M_{\text{background}}}{M_{\text{bb}} - M_{\text{background}}} \qquad (8)$$

More details of the measurements can be found in reference [13].